# A robust direct modeling method for quadric B-rep models based on geometry-topology inconsistency tracking

Qiang Zou*, Hsi-Yung Feng
The University of British Columbia
Vancouver, BC, Canada V6T 1Z4

**Abstract**

Boundary representation (B-rep) model editing plays an essential role in computer-aided design and motivates the very recent direct modeling paradigm, which features intuitive push-pull manipulation of the model geometry. In mechanical design, a substantial part of B-rep models being used are quadric models (composed of linear and quadric surfaces). However, push-pulling such models is not trivial due to the possible smooth face-face connections in the models. The major issue is that, during push-pull moves, it is often desirable to preserve these connections for functional, manufacturing, or aesthetic reasons, but this could cause complex inconsistencies between the geometry and topology in the model and lead to robustness issues in updating the model. The challenge lies in effectiveness towards detecting the instants when geometry-topology inconsistencies occur during push-pull moves. This paper proposes a novel reverse detection method to solve the challenge and then, based on it, presents a robust method for push-pull direct modeling while preserving smooth connections. Case studies and comparisons have been conducted to demonstrate the effectiveness of the method.

**1. Introduction**

Boundary representation (B-rep) models are fundamental to computer-aided design (CAD). Particularly, almost all export/import 3D CAD models are B-rep solid models made in the STEP/IGES format. Interactive environments for editing such models thus play an essential part in modern CAD systems. These environments require intuitive, flexible model manipulation and fast model update. The very recent direct modeling CAD paradigm encompasses all these characteristics. In direct modeling, the primary feature/function is the intuitive push-pull interaction with the geometry (boundary faces) of a model [1,2]. Such push-pull controls allow models to be modified very easily, and ultimately leading to faster iterations between design alternatives and simulations.

Push-pull direct modeling allows users to effectively edit a solid model through grabbing, pushing, and pulling its boundary faces, refer to [3] for some examples. On the other hand, a push-pulled boundary face could cause changes made to the connections between this face and its neighboring boundary faces. If the changes are intended exactly by the user or the changed connections are trivial, this way of working is more than adequate. Nevertheless, there are also many scenarios in which these boundary faces are connected in a smooth manner for considerations like stress concentration reduction, manufacturability, injury prevention (from sharp edges), or even aesthetic design [4], which makes the connections non-trivial. In view of these design intents, the user may want the smooth connections to be preserved during push-pull moves. It is thus necessary for direct modelers to provide the option of push-pull direct modeling while preserving smooth connections.

---

* Corresponding author.  E-mail addresses: john.qiangzou@gmail.com (Q. Zou); feng@mech.ubc.ca (H-Y Feng)

The smooth connections can be $G^1$ or higher. $G^1$ is the first order geometric continuity and states that two faces share colinear normal directions along their joint edge (or equivalently, they are connected tangentially) [5]. Smooth connections of a higher order than $G^1$ are also seen in B-rep models but only in those involving free-form surfaces. This work focuses on quadric models that are composed of linear and quadric surfaces, and such models cover about 95% of models in mechanical design [6]. In this regard, the problem to be studied in this work is as follows: push-pull direct modeling of featureless quadric B-rep solid models while preserving the involved $G^1$ continuous connections (shortly, push-pull with $G^1$ connections).

Push-pull with $G^1$ connections is no trivial matter. A B-rep model consists of information on both geometry and topology, which must be consistent with each other to attain a valid solid model [7]. When a valid model is push-pulled, the moved boundary faces will cause changes made to the model's geometry and topology, which in turn could break the information consistency in the model [3]. (Fig. 1 shows one such example.) To successfully resolve these inconsistencies, knowing the instants when they occur is important [8]. Missing any such instant would leave some inconsistencies unaddressed, and consequently the direct modeler gives model update failures and exhibits robustness issues. Detecting geometry-topology inconsistencies (GTI) is not an easy task and becomes particularly challenging when it comes to push-pull with $G^1$ connections because, to preserve $G^1$ connections, neighboring faces of push-pulled faces need to be made movable, but their motions are not quite known in fact. Without knowing the motion information, it is clearly difficult to detect GTI effectively.

To be more specific on the above challenge, a closer look at movable boundary faces needs to be made. There are two types of movable boundary faces in push-pull with $G^1$ connections: (1) the boundary faces push-pulled by the user, and (2) the neighboring boundary faces[1] driven by the push-pulled boundary faces (to keep the $G^1$ connections). The driven boundary faces move according to how the user moves the push-pulled boundary faces, and their motions are essentially governed by a system of tangent constraints representing the $G^1$ connections. Not until this system is solved do we know positions/orientations of the driven boundary faces for any intermediate instants during the push-pull move. There are thus no explicit expressions for the driven boundary faces' motions. One may obtain an approximation to the motions through a brute-force sampling/solving, but the resulting high computational load makes this strategy unattractive. As a result, there is a lack of the driven boundary faces' motion information in push-pull with $G^1$ connections; GTI detection then becomes a challenging problem.

Although some attempts about GTI detection have been made in the literature, their applicability is consistently limited to push-pull edits without considering smooth connections. The solutions provided by industry are not satisfactory either; push-pull with $G^1$ connections is either partially supported for a few selected scenarios or fully supported but with robustness issues (as will be shown in Section 4). New developments are thus necessary to have effective GTI detection and robust push-pull with $G^1$ connections. This work presents a novel method for GTI detection: we typically miss instants of GTI first, and then reversely catch the instants after they have actually happened, a posteriori. This reverse method allows for effective GTI detection while avoiding reliance on motion information of movable boundary faces. With it, a robust method for push-pull with $G^1$ connections can be developed.

---

[1] Their neighboring boundary faces may also be included if necessary, and so forth.

## 2. Related work

The notion of push-pull direct modeling was initially proposed by industry to meet the increasing need for efficient and flexible model modification in design reuse. Push-pull with $G^1$ connections has been implemented in a few commercial CAD systems, but the implementation information is kept private or patented, as it may lead to competitive advantages for CAD vendors. As a result, none of the CAD systems comes with guarantees or clearly stated limitations. Robust issues are observed in these CAD systems: they work well in some scenarios but not in others. In some CAD systems, push-pull with $G^1$ connections is not even fully supported. For example, Autodesk Inventor is only able to preserve $G^1$ connections relating to fillets. In this regard, the current solutions provided by industry are far from being complete.

In the literature, there is a limited number of publications related to push-pull direct modeling since it is a relatively new notion in CAD. Lipp et al. [8] presented a GTI detection method but restricted models to solid polygonal meshes that are composed only of planar faces; such models do not involve $G^1$ connections. Zou and Feng [3] proposed a continuity-based method for push-pulling solid models, but no particular attention was given to smooth connections. In both methods, heuristics were employed to detect GTI, and knowing beforehand the motion information of all movable boundary faces is a necessary condition for the heuristics to work properly. This makes their methods inapplicable to the problem considered in this work. There are also studies [9–12] approaching push-pull direct modeling from the perspective of feature-based modeling. These methods essentially translate push-pull edits to parametric feature edits. Such a strategy is clearly not suitable for handling featureless B-rep solid models as in this work.

In parametric modeling, there is a collection of studies [13–15] that are related but not directly connected to the present work. Their interest is to compute the parameter range within which the model topology remains unchanged under parametric edits [13]. The limits of this parameter range are the critical points at which the model topology changes. The notion of these critical points is conceptually related to the GTI detection task in this work. However, the methods presented in Refs. [13–15] are not suitable for this work due to the reliance on parametric relationships in the model, which are not available in featureless B-rep models.

The above review suggests that documented studies and the industrial state of the art on the problem of push-pull with $G^1$ connections are quite insufficient. A new method is to be presented to address this insufficiency in the following Section 3, in particular on effective GTI detection. Section 4 will validate the proposed method with a series of case studies, as well as illustrating its limitations. Conclusions of the paper are given in Section 5.

## 3. The proposed methodology

For better presenting the proposed method, several important notions to be used in the following text are made precise first. A *Push-pull* is the edit of a model by translating and/or rotating its boundary faces. Every push-pull edit is followed by a regeneration of the boundary representation for the push-pulled model, referred to as *model regeneration*. This regeneration is essentially a boundary evaluation based on the post-edit carrier surfaces and the pre-edit topology [3]. Fig. 1 shows a model regeneration example with various rotational push-pulls applied to the blue face. (Here and later, blue faces in figures indicate push-pulled faces.) The aforementioned *GTI* refers to the situation in which the regenerated model possesses more connections than those in the pre-edit topology, as shown by the newly inserted connection in Fig. 1; or the other way around, it is impossible to form some connections in the pre-edit topology regarding the configuration of the post-edit carrier surfaces. (The interested reader is referred to

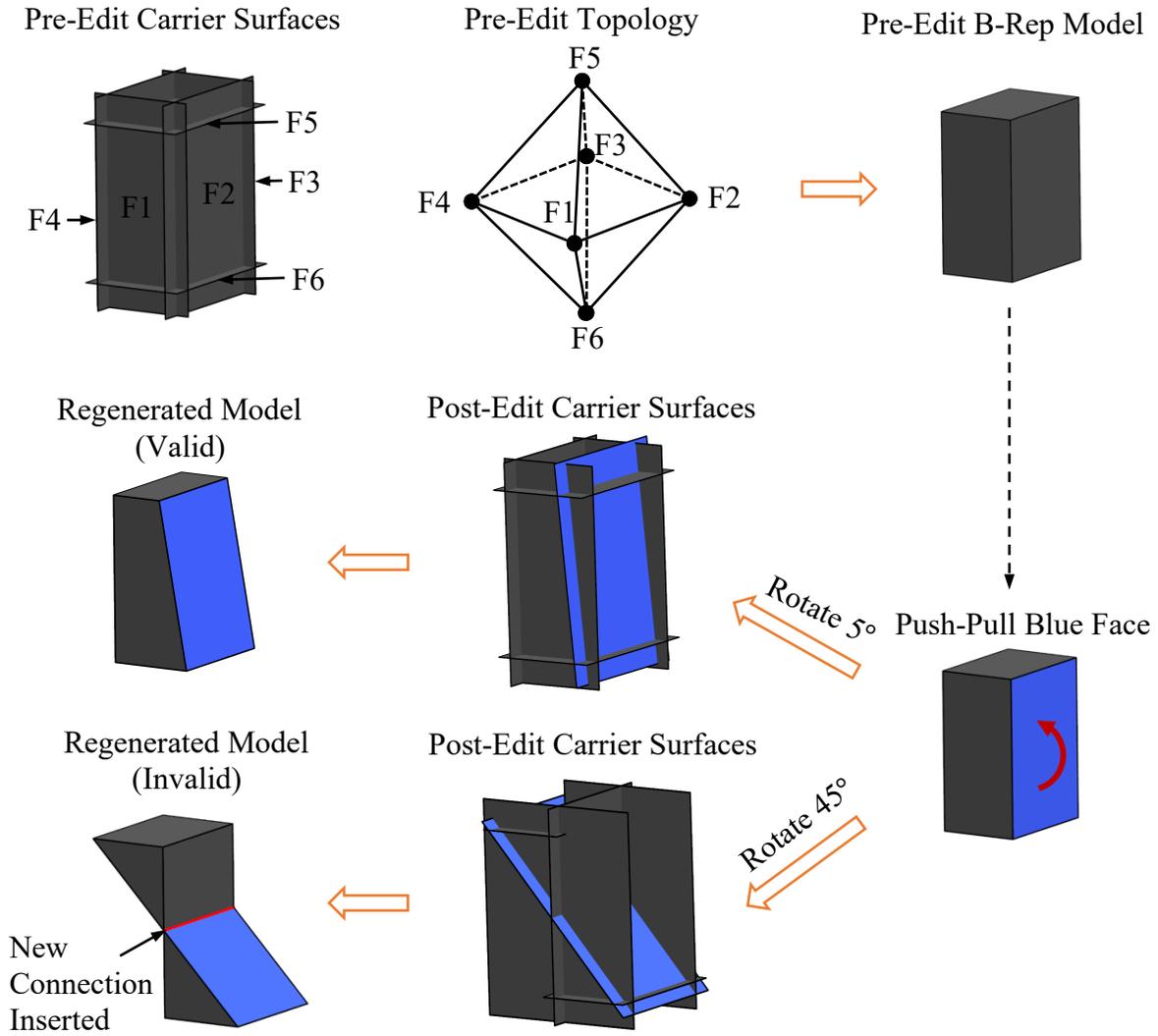

Figure 1: Illustration of model regeneration and geometry-topology inconsistency.

Section 3, Ref. [3] for a detailed discussion of GTI.) If there is any GTI, model regeneration fails to output a valid solid model, necessitating *model update* that resolves the GTI to attain a valid modeling result.

The overall method presented in this work is shown in Fig. 2. The part in the dashed rectangle represents the main body of the method, which iteratively applies two procedures: GTI detection and GTI resolution. GTI detection is to evaluate the next critical point at which GTI occurs during a push-pull edit of interest; GTI resolution is to resolve the GTI immediately after it is detected. The resulting model will serve as the base model in the next iteration for carrying out the rest part of the push-pull edit. These two procedures are repeated until no more critical point can be detected. Model regeneration then gives the intended model for the whole push-pull edit. It should be noted that this iterative strategy is not new; it has been used in the previous work [3,8]. Nevertheless, as already noted, the previously developed GTI detection methods do not apply to the problem of push-pull with $G^1$ connections. Without an effective GTI detection method, the above iterative strategy cannot be made possible. This issue is to be addressed in Section 3.1. For the GTI resolution module, the methods developed in [3,8] are still applicable since push-pull with $G^1$ connections do not introduce additional inconsistency types. Thus, a similar method is to be used for the GTI resolution task in this work, as will be shown in Section 3.2.

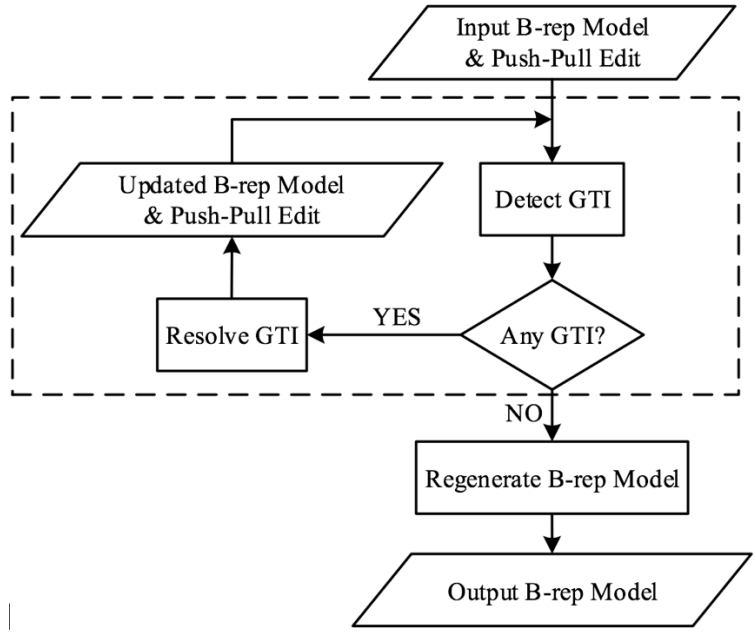

Figure 2: Schematic diagram of the overall method.

### 3.1. Geometry-topology inconsistency detection

Although presented in different forms, previous studies on GTI detection, e.g., [3,8,14,15], share a common idea: they predict ahead of time when GTI will occur during a push-pull edit, based on various GTI pattern libraries. Basically, GTI detection was done a priori, and fully knowing motion information of all movable boundary faces is a necessity. Such a strategy will suffer if the motion information is unavailable. For this reason, this work does not seek to improve existing methods but to propose a radically different strategy. The basic idea is to detect GTI a posteriori: we first miss instants of GTI, then investigate the generated GTI in the regenerated model, and finally reversely trackback to the critical points where the GTI occurred, a concept similar to reverse engineering. This way, the motion information does not need to be made available ahead of time, and then the challenge stated in Section 1 can be properly solved. The above idea is to be referred to as the reverse inference idea.

The reverse inference idea is implemented using a three-stage scheme, as illustrated in Fig. 3. An example based on a mechanical connecting rod part is also given to assist the understanding of the workflow. At Stage 1, a model regeneration is first performed. If there is no GTI in the regenerated model (i.e., a valid model is produced), nothing further needs to be done. If otherwise, the GTI in the regenerated model is extracted. As already noted, GTI takes the form of newly added connections and/or lost connections. Such connection changes are reflected in the changes made to the edge topology on the boundary faces [3], as connections essentially define intersections between carrier surfaces, as shown by the circled face in Fig. 3.[2] Then, the extraction can be done by collecting topology-changed boundary faces in the regenerated model.

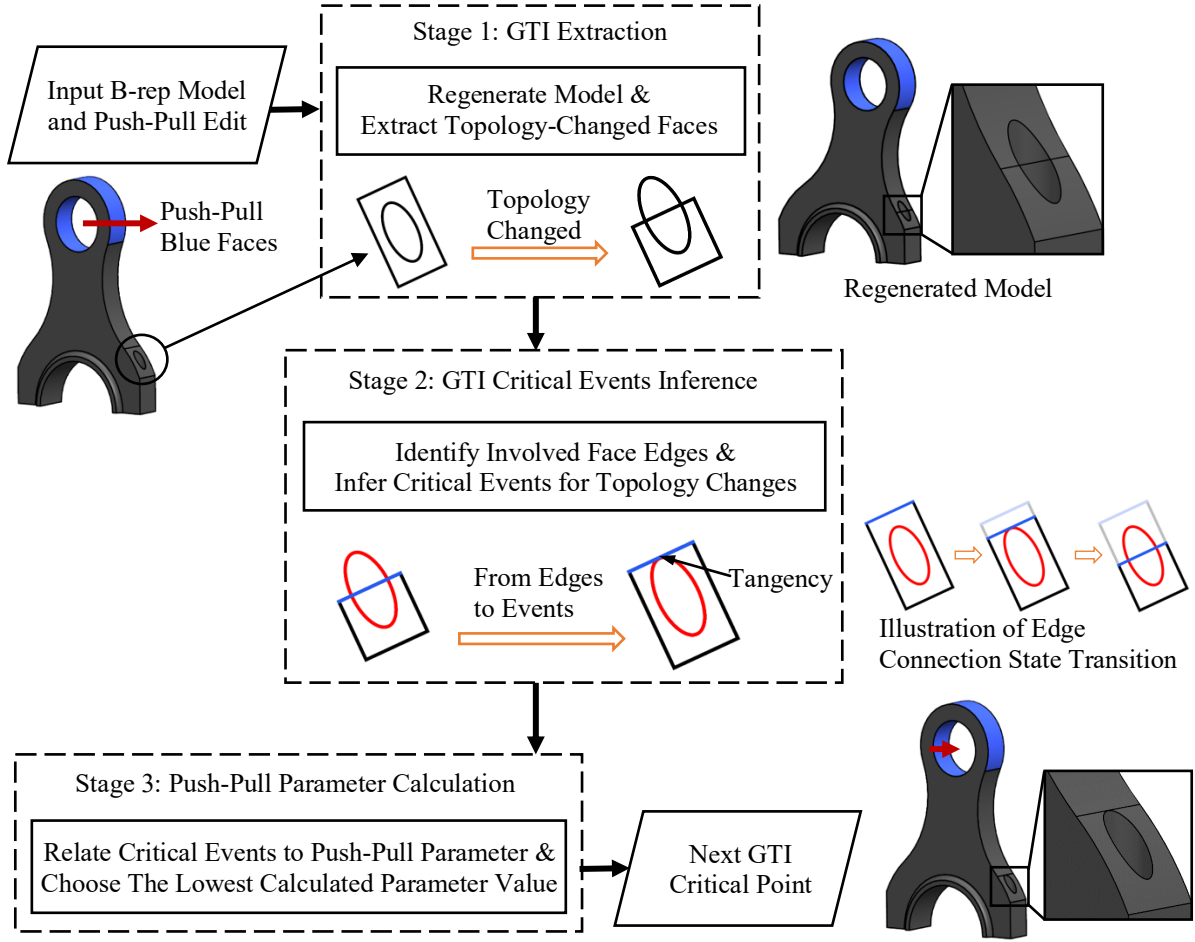

Figure 3: Workflow of geometry-topology inconsistency detection.

Stage 2 investigates how the topology changes on individual ill-bounded faces collected at Stage 1. This is done by first identifying the edges involved in the changed topology (as shown by the blue and red edges in Fig. 3) and then analyzing how their connection states alter: how two edges change from being apart to being connected, or the other way around. This analysis serves as a basis for reversely inferring the critical events where the edge connection states change abruptly. For example, the critical event for the blue and red edges in Fig. 3 is the tangency between them, as depicted in the transition illustration. Inferring such events may be straightforward using human intuition but not the case with a computer. In the following, an effective method will be presented to handle this issue.

At Stage 3, we model the relationship between the push-pull parameter and the critical events attained at Stage 2. A push-pull operation can be represented by the rigid transformation matrix $T(t), t \in [0,1]$ specified by the user. The parameter $t$ here is the push-pull parameter, and the parameter domain has been normalized into the unit interval [0,1]. Each critical event has a corresponding push-pull parameter value (called a critical point) indicating when the critical event occurred, and it is to be shown that this correspondence can be described by a system of nonlinear equations. Solving this system for individual critical events gives a collective list of critical points, and the lowest value of these critical points is the intended GTI detection result.

---

[2] In implementation, one may confront a very special situation in model regeneration: number of intersection for two surfaces may change, and choices among the intersections need to be made. Handling this issue is tedious and beyond the scope of this work. Please refer to the persistent naming literature for detailed discussions and possible solutions.

Among the three stages stated above, the essential parts are the reverse inferring from topology-changed faces to critical events and the relating of these events to the push-pull parameter. Novel methods are to be presented to deal with them in the next few subsections.

### 3.1.1. Reverse inferring of critical events

This subsection addresses the problem of attaining critical events that characterize abrupt topology changes on given topology-changed boundary faces. A topology-changed face is a face in the regenerated model, and thus to be called the regenerated face; its corresponding face in the pre-edit model is to be referred to as the reference face. Recall that the very first task in the reverse inference is to attain the edges involved in topology changes, which gives rise to the need for retrieving and manipulating the topological information of a boundary face. To facilitate such processing, a graph-based face representation scheme is used: connections of bounding edges on a boundary face are represented by a graph structure in which graph nodes encode the edge entities and graph arcs describe the connections between them. Fig. 4 shows an example of this graph representation scheme, using the reference face and regenerated face of the connecting rod example in Fig. 3.

The primary advantage of using the graph representation is the many operators available for manipulating graphs. Those of interest are the Boolean operations that can extract the difference between the two graphs. To be more specific, the difference between the reference and regenerated faces' connection graphs $G_{ref}$, $G_{reg}$ is given by:

$$\Delta G = (G_{reg} - G_{ref}) \cup (G_{ref} - G_{reg}) \qquad (1)$$

where the operators "−" and "∪" denote the Boolean subtraction and Boolean union on graphs. Boolean operations on graphs have various definitions [16], and the one used here is as follows. Let two graphs be $G_1 = (V_1, A_1)$ and $G_2 = (V_2, A_2)$, where $V_1$ and $V_2$ are the graphs' node sets, and $A_1$ and $A_2$ are the graphs' arc sets. The Boolean subtraction is then given by $G_1 - G_2 = (\{\text{nodes induced from } A_1 - A_2\}, A_1 - A_2)$, where the operator "−" is the ordinary set difference; the Boolean union can be defined similarly. Following this definition, the first subtraction $G_{reg} - G_{ref}$ in Eq. (1) acquires arcs presented in $G_{reg}$ but not in $G_{ref}$. As arcs represent connections between edges entities on a face, $G_{reg} - G_{ref}$ essentially describes connections that $G_{reg}$ has but $G_{ref}$ does not; such connections are referred to as newly added connections. Similarly, $G_{ref} - G_{reg}$ gives connections that $G_{ref}$ has but $G_{reg}$ does not, referred to as lost connections. Then the union of these two terms yields the difference between $G_{reg}$ and $G_{ref}$ in terms of connections. For example, applying Eq. (1) to the two connection graphs in Fig. 4 yields a difference graph that correctly captures the newly added connection between edges $e_1$ and $e_5$.

The difference graph in Eq. (1) not only tells the subjects of topology changes but also carries information on the progression history of topology changes. Given a difference graph $\Delta G = (\Delta V, \Delta A)$, we can easily check if an arc $a \in \Delta A$ is from the regenerated connection graph $G_{reg}$ or the reference connection graph $G_{ref}$. If $a$ is from $G_{reg}$, it represents an newly added connection, and the two edges' connection state progressed from being apart to being connected. If $a$ is from $G_{ref}$, it represents an lost connection, and the connection state progressed from being connected to being apart. Consider again the example in Fig. 4. The only arc of the difference graph comes from the connection graph B (i.e., $G_{reg}$), and then the topology change is a consequence of edges $e_1$ and $e_5$ moving into each other, as illustrated by the red arrow.

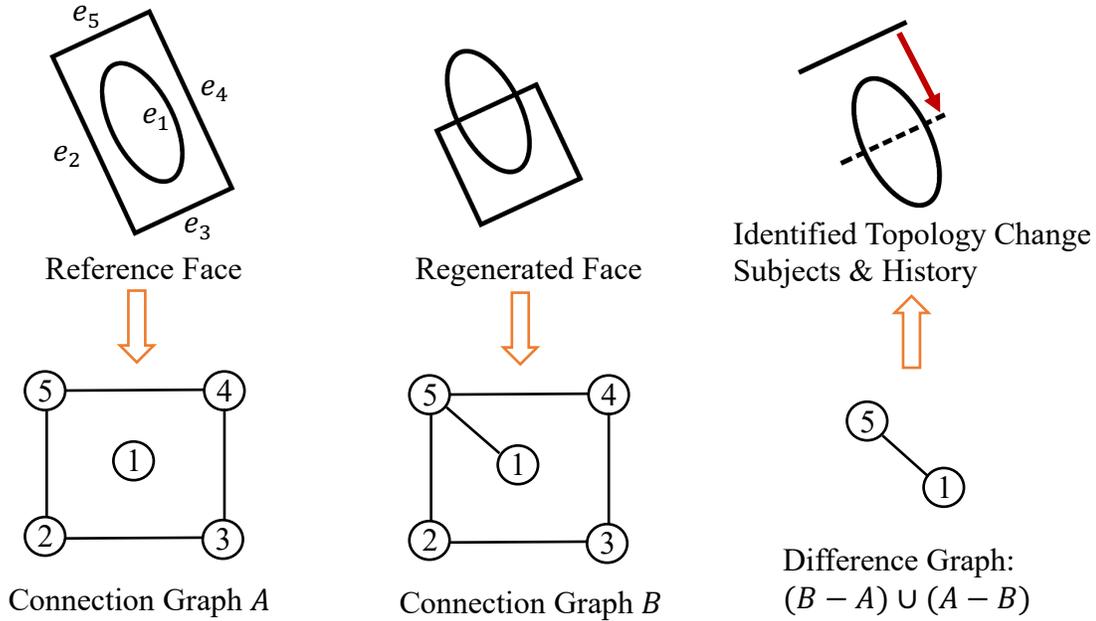

Figure 4: Graph representation of boundary faces and identification of topology change subjects and history.

A difference graph $\Delta G$ gives a list of edge pairs (having connection state changes) and their individual progression histories. Each progression thread makes either two originally disconnected edges connected, or the other way around. The critical event for the progression from disconnection to connection is the collision between the two edges. Generally, edge-edge collision has two configurations (1) involving only interior points of both edges and (2) having end points involved, as shown in Fig. 5. These two configurations have different characteristic events. As can be seen from Fig. 5, the event for the left configuration is the interior-point tangency between the two edges, and that for the right one is the end-point incidence. It should be noted here that no further information is available for us to determine which of the two critical events has actually occurred since we do not have the motion information of all movable boundary faces. As a result, both events are possible and need to be checked. For the second progression type, i.e., from connection to disconnection, the critical event is the separation between the carrier curves of the two edges, as shown in Fig. 6. More precisely, one of the curves moved beyond the size limit of the other curve, and the characteristic event is the tangency at the ends of the two edges.

*3.1.2. Mathematical modeling of critical events*

With critical events in place, we need to relate them to the push-pull parameter so as to attain the critical points at which these events occur. From the previous discussion, there are three critical events: interior-point tangency, end-point incidence, and end-point tangency. They can be further reduced to the following two basic types: tangency and incidence. It will be shown in the following that these two event types can be formulated as systems of quadric equations.

Topology Change Tracking in Direct Modeling

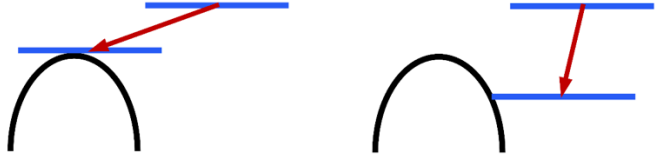

Figure 5: Example configurations of edge-edge collision.

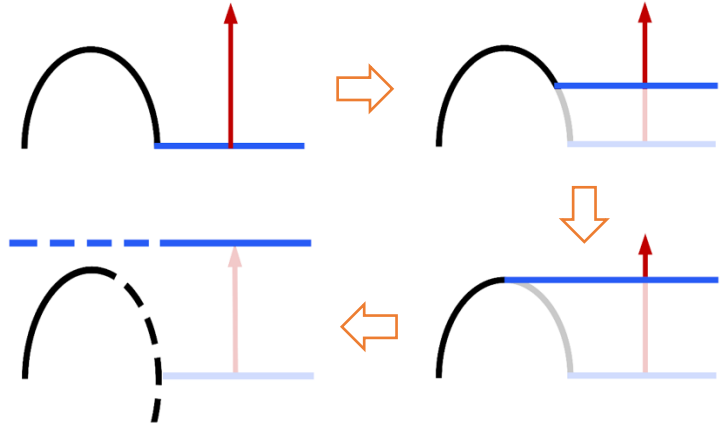

Figure 6: Example process of edge-edge separation.

A tangency (interior-point or end-point) between two edges can be expressed sequentially in terms of (1) the tangency between the two edges' carrier curves and (2) the examination of the tangency locating inside or right at the limits of the edges. More formally, let the edges be represented as $e_1, e_2$, their respective carrier curves be $c_1, c_2$, and assume that the tangency of the two curves is at point $p$. Then, $p$ clearly has to satisfy the on-curve conditions:

$$p \in c_1 \text{ and } p \in c_2 \qquad (2)$$

In general, carrier curves in a B-rep model are intersections of surfaces [17]. Thus, the on-curve conditions can be modeled as:

$$p \in c_1 \Leftrightarrow \begin{cases} F_1(p) = 0 \\ F_2(p) = 0 \end{cases}$$
$$p \in c_2 \Leftrightarrow \begin{cases} F_2(p) = 0 \\ F_3(p) = 0 \end{cases} \qquad (3)$$

where $F_1(\cdot) = 0$, $F_2(\cdot) = 0$ and $F_3(\cdot) = 0$ denote the equations of the surfaces adjacent to carrier curves $c_1, c_2$, refer to Fig. 7 for an illustration of their relationship.

In addition to the on-curve conditions, point $p$ also needs to satisfy the tangency condition that the two curves have collinear tangents at point $p$. The tangent for curve $c_1$ (or $c_2$) is given by the cross product of the normals of the surfaces $F_1$ and $F_2$ (or, $F_2$ and $F_3$). The normal of a surface $F(x, y, z) = 0$ is given by its gradient: $\nabla F = \left(\frac{\partial F}{\partial x}, \frac{\partial F}{\partial Y}, \frac{\partial F}{\partial Z}\right)$ [18]. The tangency condition can then be formulated as:

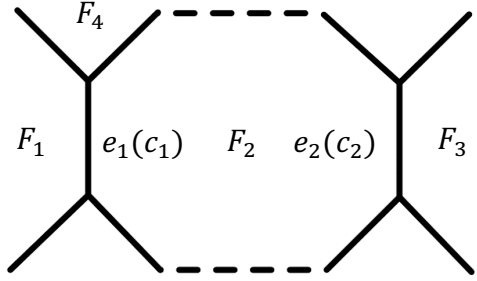

Figure 7: Illustration of edge-curve-surface relationship.

$$\begin{gathered}(\nabla F_1(p) \times \nabla F_2(p)) \times (\nabla F_2(p) \times \nabla F_3(p)) = \mathbf{0} \\ \Leftrightarrow \\ \nabla F_2(p) \cdot (\nabla F_1(p) \times \nabla F_3(p)) = 0\end{gathered} \quad (4)$$

where operators "×" and "·" denote the cross product and dot product, respectively. Combining Eq. (4) with Eq. (3) yields the mathematical modeling of tangency events.

For an incidence event at edge ends, it should first satisfy the on-curve conditions (Eq. (3)) as well, and then satisfy the additional constraint that point $p$ is at edge ends. As shown in Fig. 7, the top end of edge $e_1$ is the intersection of curve $c_1$ and the surface $F_4$. Thus, to impose the edge end constraint, we only need to require that point $p$ is on surface $F_4$, which is given by:

$$F_4(p) = 0 \quad (5)$$

This equation, together with Eq. (3), gives the mathematical modeling of incidence events.

In the above formulations, there are four equations but three variables (i.e., the three coordinates of point $p$) for the tangency event (Eq. (3) and (4)), and the same for the incidence event (Eq. (3) and (5)). The missing variable is reserved for the push-pull parameter. Let the push-pull edit be represented by a rigid transformation matrix $T(t), t \in [0,1]$. This transformation matrix imposes the motion on push-pulled boundary faces, which in turn drive the neighboring boundary faces to move due to the $G^1$ connections between them, and so forth for their neighboring boundary faces if necessary. The motions of all the driven boundary faces are governed by a system of tangent constraints between their carrier surfaces, which ensures that the $G^1$ connections between the boundary faces are preserved during the push-pull move. These tangent constraints can be translated straightforwardly to a system of nonlinear equations using the existing research results in geometric constraint solving such as [19–21]. Combining these equations with Eq. (3) and (4) (or (5)) yields a new system of nonlinear equations that relate the push-pull parameter to a critical event previously detected. In other words, by solving this new system, we can attain the exact value (up to a given precision) of the parameter at which a critical event occurs.

To summarize, Algorithm 1 shows how to combine the methods described previously to attain an effective GTI detection method. This algorithm follows the workflow presented in Fig. 3: first attain topology-changed boundary faces (Line 2); then collect critical events using the method described in Section 3.1.1 (Line 5); then relate the collected critical events to the push-pull parameter with the method described in Section 3.1.2 (Lines 6-9); finally set the next critical point to the lowest value of all the calculated critical points.

Topology Change Tracking in Direct Modeling

---

**Algorithm 1**: Geometry-Topology Inconsistency Detection
**Input:** $M, T(t), t \in [0,1]$ — the B-rep model and push-pull edit
**Output:** the next critical point

1. $M' \leftarrow \text{RegenerateModel}(M, T(t=1))$
2. $F \leftarrow \text{GetTopologyChangedFaces}(M')$
3. $T' \leftarrow \emptyset$  // for storing critical points
4. **for** each face $f \in F$ **do**
5.     $E \leftarrow \text{InferCriticalEvents}(f)$
6.     **for** each event $e \in E$ **do**
7.         $t' \leftarrow \text{RelateEventToPushPullParameter}(e, t)$  // Eq. (3), (4) and (5)
8.         add $t'$ to $T'$ if $0 \leq t' \leq 1$  // filter out invalid critical points
9.     **end for**
10. **end for**
11. **Return** $\text{Min}(T')$

---

### 3.2. Geometry-topology inconsistency resolution

As can be seen from the workflow in Fig. 2, the subsequent task after attaining the next critical point is to resolve the GTI generated when the push-pull edit crosses this critical point. The main task is to ensure that the model after resolution remains as a valid solid model. The authors have previously presented a Boolean-based GTI resolution method to address this problem [3]. This method can ensure valid modeling results and attain continuous model variations. It is summarized below.

Let the push-pull edit be represented by a rigid transformation matrix $T(t), t \in [0,1]$, and the model variation over this push-pull move by $M(t), t \in [0,1]$. The next critical point is assumed to be at $0 < \tau < 1$. As there is no GTI at any point before $\tau$, the model $M(t)$ is simply the regenerated model at $t$, for $0 < t < \tau$. For such a model variation, it is found that the model volume change $\Delta M(t) = M(0) - M(t), t \in (0, \tau)$ can be expressed in terms of the volume swept by the push-pulled face from 0 to $t$ and bounded by its neighboring faces. Here, the operator "−" denotes the regularized Boolean difference on the model volume. Then, the model $M(t)$ before the critical point $\tau$ can be obtained by a simple Boolean difference operation:

$$M(t) = M(0) - \Delta M(t), \quad t \in [0, \tau) \qquad (6)$$

In the above expression, it is evident that if $\Delta M(t)$ is negative, material/volume will be added to the original model $M(0)$.

As the model variation is required to follow a continuous change pattern, the following relationship holds for the critical point: $M(\tau) = \lim_{\epsilon \to 0} M(\tau - \epsilon)$. The model $M(\tau - \epsilon)$ can be evaluated using Eq. (6). Eventually, the model $M(\tau)$ is given by:

$$M(\tau) = M(0) - \lim_{\epsilon \to 0} \Delta M(\tau - \epsilon) = M(0) - \Delta M(\tau) \qquad (7)$$

where $\Delta M(\tau)$ is, again, the volume swept by the push-pulled face from 0 to $\tau$ and bounded by its neighboring faces. With Eq. (7), resolving the GTI at a critical point is formulated to be Boolean operations on the model volume. The above statements represent a very brief, high-level description of the Boolean-based resolution method. The method clearly involves more technical details such as addressing interferences among the model $\Delta M(t)$ for multiple moved boundary faces. These issues can be solved with variations of the idea stated above, and the details can be found in Section 5, Ref. [3].

Use of Boolean operations to carry out GTI resolution provides substantial advantages. Most notably, Boolean operations on the model volume can guarantee that the resulting model is a valid solid model [9]. In addition, they allow an easy implementation of push-pull with $G^1$ connections using existing CAD research and engineering results.

### 3.3. Overall algorithm for push-pull with $G^1$ connections

With the GTI detection method (Section 3.1) and the GTI resolution method (Section 3.2) in place, an algorithm is readily available for handling any possible GTI in push-pull with $G^1$ connections. The algorithm is essentially an implementation of the diagram in Fig. 2 by replacing the two blocks of Detect GTI and Resolve GTI with the respective methods in Sections 3.1 and 3.2. The specific procedures of the algorithm are shown in Algorithm 2. Line 2 corresponds to the inconsistency detection method described in Section 3.1. Line 4 corresponds to the Boolean-based resolution method described in Section 3.2. Line 5 updates the B-rep model with the model resulting from the inconsistency resolution, and Line 6 updates the push-pull edit with the rest part, which is given by $T(1)\,T(t')^{-1}$. These updates (Lines 5-7) prepare the model and push-pull edit for the next iteration.

---

**Algorithm 2**: Push-Pull with $G^1$ Connections
**Input:** $M, T(t), t \in [0,1]$ — the B-rep model and push-pull edit
**Output:** the updated model

1. **while** *TRUE*
2.    $t' \leftarrow$ DetectNextCriticalPoint$(M, T)$ // Algorithms 1, Section 3.1
3.    **if** $t' \neq$ *NULL* **then** // *NULL* means there is no critical point
4.      $M' \leftarrow$ ResolveInconsistency$(M, T(t'))$ // Section 3.2
5.      $M \leftarrow M'$ // update the B-rep model
6.      $T \leftarrow T(1)\,T(t')^{-1}$ // update the push-pull edit
7.      Reparametrize $T$ to the domain $[0,1]$
8.    **else** break the loop
9. **end while**
10. $M \leftarrow$ RegenerateModel$(M, T(1))$
11. **Return** $M$

---

## 4. Case studies

### 4.1. Implementation

The method presented previously has been implemented in an Apple Macintosh environment using C++. The implementation was built on top of the open source geometric modeling kernel Open CASCADE (7.0); the graphical user interface was designed using QT (5.7); the software's architecture is similar to the geometry processing and rendering framework OpenFlipper [22]. When a STEP/IGES model is imported, its boundary representation is displayed in the View window, and its handle is listed in the Objects toolbox (labeled as 1 in Fig. 8a). The handle allows the user to control the model's display parameters like hide/show. To push-pull a B-rep model displayed in the View window, the user first activates the push-pull direct modeling function by pressing the left button in the Push-Pull toolbox (labeled as 2 in Fig. 8a) and then selects the boundary faces of interest. Once done, a push-pull handle (labeled as 3-5 in Fig. 8a) pops up. By selecting and moving the three rectangular sub-handles in the push-pull handle, users can push-pull the selected boundary faces as they see fit. After the push-pull parameters are specified, the computer uses the method presented in Section 3 to carry out the model update. For instance, Fig. 8b shows the updated model for translating the blue faces by $7.5\ mm$.

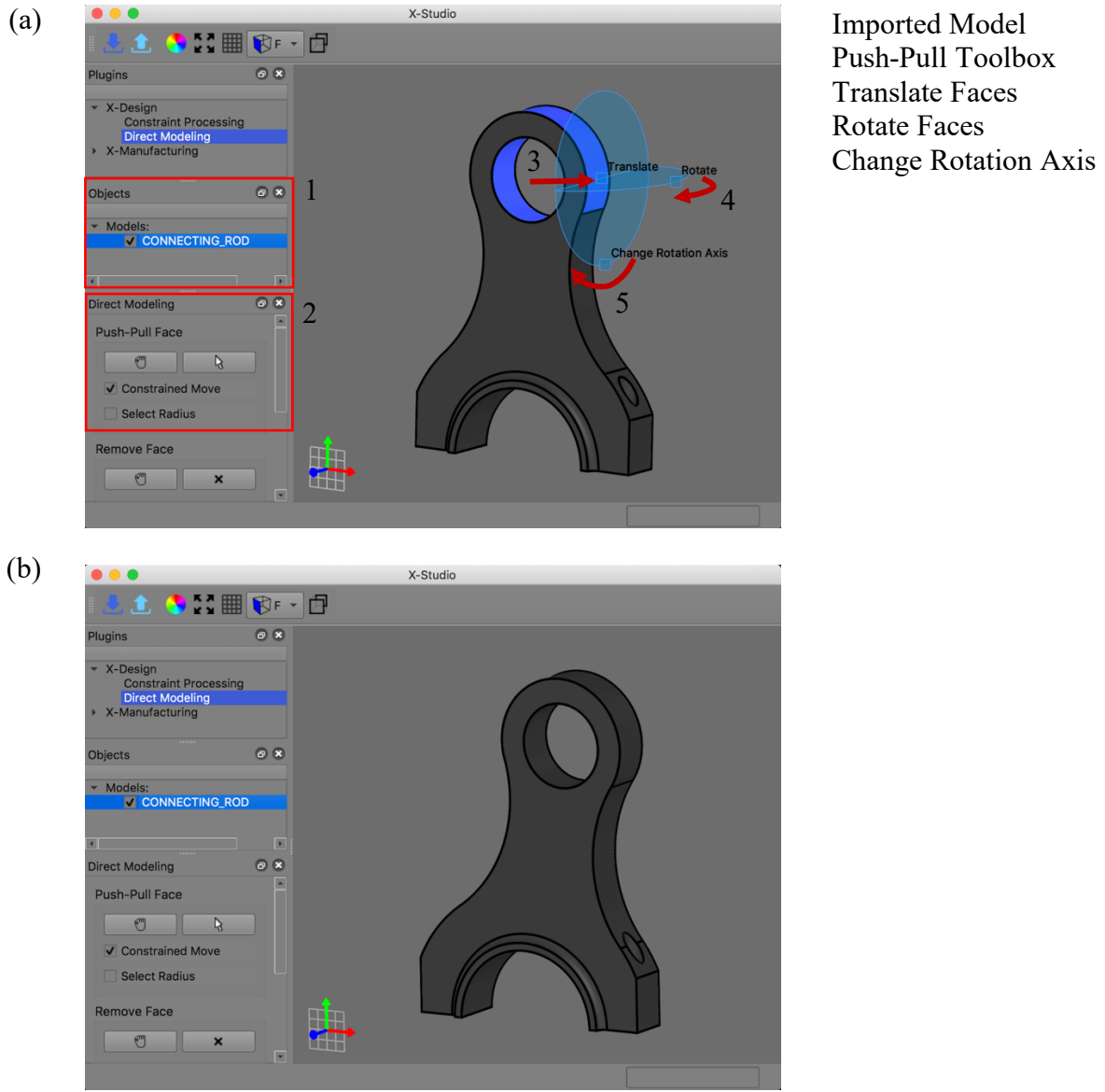

Figure 8: Graphical user interface (a) of the push-pull with $G^1$ connection system and (b) modeling result of translating blue faces.

**4.2. Case studies**

A series of case studies are to be presented to demonstrate the effectiveness of the proposed method by comparisons with commercial CAD systems. Five leading CAD systems were tested: ANSYS SpaceClaim (19), Autodesk Inventor (2019), PTC Creo (Elements/Direct modeling 19), Siemens NX (11), and SolidWorks (2019). SolidWorks barely supports push-pull with $G^1$ connections; Autodesk Inventor and PTC Creo partially support push-pull with $G^1$ connections in a few selected scenarios; ANSYS SpaceClaim and Siemens NX fully support push-pull with $G^1$ connections. Between ANSYS SpaceClaim and Siemens NX, the latter has a better performance in terms of preserving $G^1$ connections; sometimes the former even gives nonsense model shapes. (ANSYS SpaceClaim, however, outperforms Siemens NX

Topology Change Tracking in Direct Modeling

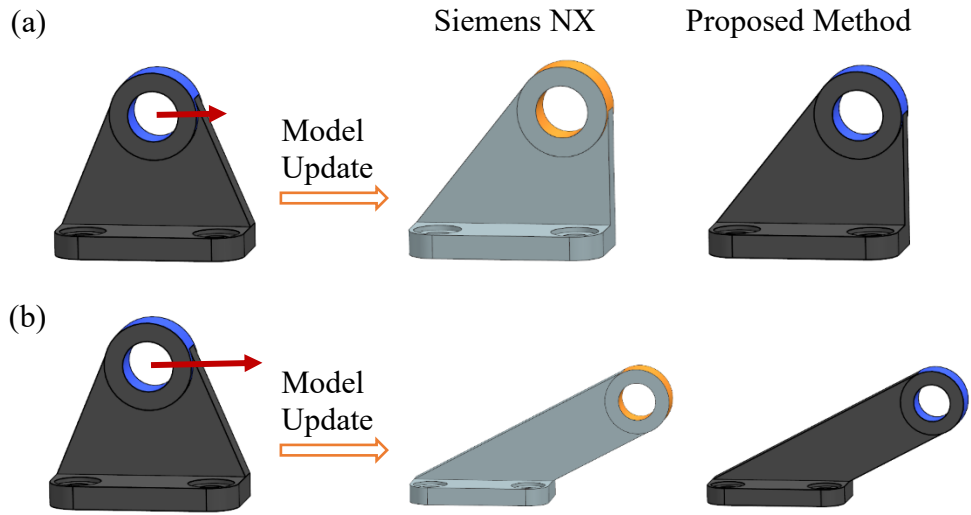

Figure 9: Push-pulling the axis support model and the modeling results: (a) an ordinary push-pull edit; and (b) an arbitrary push-pull edit.

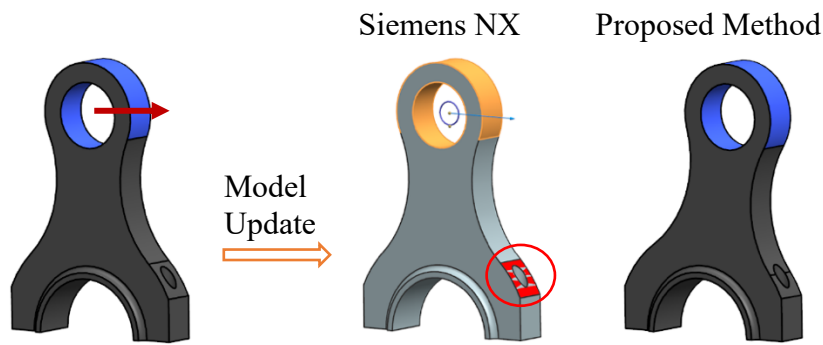

Figure 10: Push-pulling the connecting rod model and the modeling results.

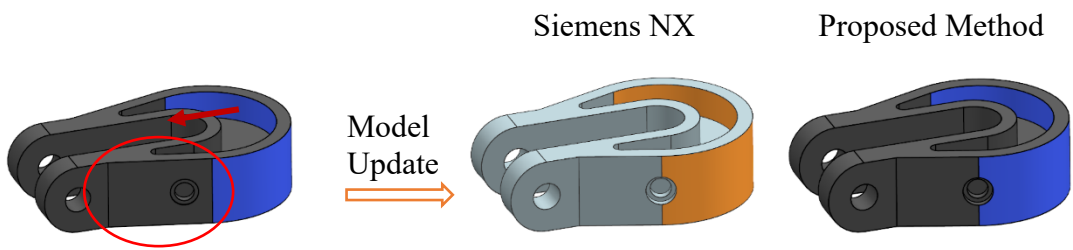

Figure 11: Push-pulling the hook model and the modeling results.

for push-pull without $G^1$ connections.) Therefore, Siemens NX was chosen for demonstrating the comparisons.

There are in total seven case studies, which are based on five real-world mechanical parts obtained from the GrabCAD part library (https://grabcad.com/library). Case study 1 considered push-pulling an axis support part model where no critical points could be detected during the push-pull edit (Fig. 9). Case study 2 involved a connecting rod part model with one critical point in the push-pull edit (Fig. 10). Case study 3 analyzed push-pulling a hook part model with one critical point in the push-pull edit (Fig. 11). The varied modeling results in these three case studies will be used to show the essential role GTI detection plays in attaining robust push-pull with $G^1$ connections, as well as the effectiveness of the proposed GTI detection method. Case studies 4-6 considered three comprehensive modeling examples where various push-pull edits were applied (Figs. 12-14); they will be used to show the effectiveness of the proposed method as a whole. Case study 7 (Fig. 15) is intended to show some important limitations of the proposed method.

### 4.3. Discussion and limitations

In case study 1, there was no critical point in the push-pull edit. Model update was thus made trivial and only involved model regeneration. Both Siemens NX and the proposed method successfully gave satisfactory modeling results (Fig. 9), even when the push-pull was made wild (Fig. 9b). In case study 2, similar faces to those in case study 1 were push-pulled, but an invalid modeling result was generated in Siemens NX (Fig. 10). (Siemens NX colors boundary faces in red whenever there is a model update failure, as shown by the circled face in Fig. 10.) The major difference between case studies 1 and 2 is that the latter involves a critical point of GTI. It can thus be concluded that crossing critical points could cause model update failures.

In case study 3 (Fig. 11), there was also one critical point of GTI, and the GTI configurations were almost the same as those in case study 2. Despite the similarity, Siemens NX failed in case study 2 and succeeded in case study 3, thereby leading to the conclusion that the failure in case study 2 was likely not due to GTI resolution but GTI detection. The failure was not likely due to numerical instability [23] either, because the geometric configurations in the two cases are also very similar. As such, the significance of GTI detection in attaining robust push-pull with $G^1$ connections can be partly confirmed. Besides, by comparing the modeling results of Siemens NX and the proposed method, the proposed GTI detection method is seen to be effective.

Case studies 1-3 focused primarily on the GTI detection module, and thus the effectiveness of the presented method as a whole was not sufficiently demonstrated. For this reason, three more case studies that are comprehensive were carried out. In case study 4 (Fig. 12), push-pulling the triple clamp model involved multiple critical points (ten in total); some of them occurred concurrently. As a result, the GTI configurations are very complex, and the associated detection task is challenging. Siemens NX was only able to successfully cross the first two critical points, while the proposed method can correctly detect all the critical points and resolve the generated GTI. In case study 5 (Fig. 13), instead of linearly adding more critical points to the push-pull move, the comprehensiveness was attained by push-puling a same model under various situations: (1) push-pull the blue faces and stop in between the first and second critical points; (2) push-pull the blue faces, and stop in between, then continue the push-pull until the end; and (3) push-pull the blue faces until the end. The modeling result for the first situation is shown in the middle of the upper row in Fig. 13, that for the second situation is shown in the upper-right, and that for the third situation is the same as the second one. Siemens NX failed to update the model for all of the three situations, while the proposed method can successfully update the model for all of them.

The comparisons in case studies 1-5 are sufficient to show that the proposed method outperforms the state of the art in terms of robustness. Nevertheless, only translational push-pulls were used in these case studies. Another case study is thus needed to show that this work also applies to other common push-pull edits in direct modeling, primarily including multiple push-pulls in a row and different push-pull types (translational and rotational). Fig. 14 shows one such example. (Noted that, as the intention here has changed, comparisons with Siemens NX are not necessary.) Four push-pull edits were performed in a row, containing both the rotational push-pull type (e.g., the first push-pull operation) and the translational push-pull type (e.g., the third push-pull operation), as well as single-face push-pull operations (e.g., the second push-pull operation) and multiple-face push-pull operations (e.g., the fourth push-pull operation). This case study represents a very comprehensive push-pull situation.

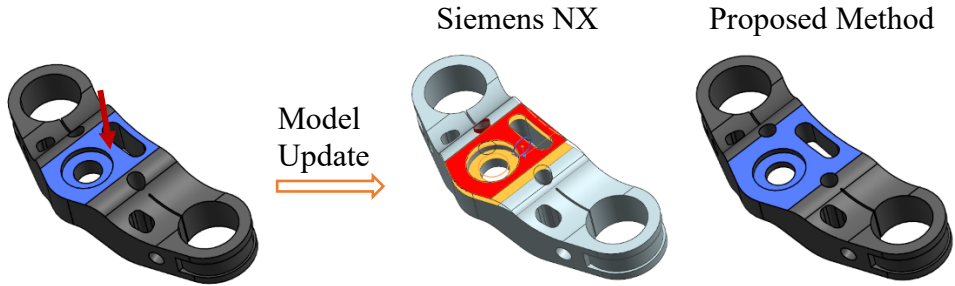

Figure 12: Push-pulling the motorcycle triple clamp model and the modeling results.

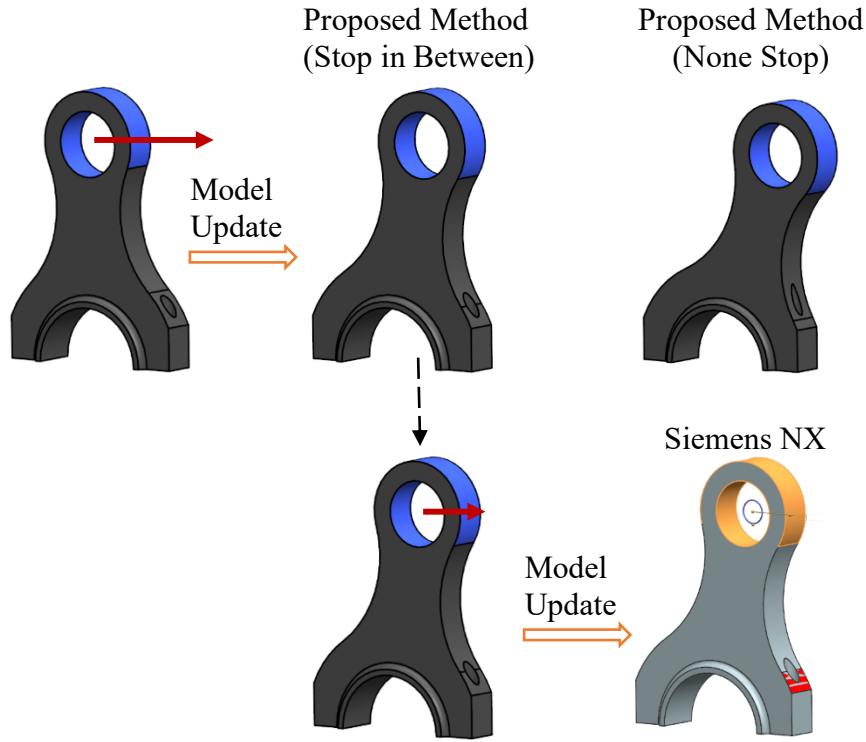

Figure 13: Push-pulling the motorcycle triple clamp model and the modeling results

As one may have already noticed, driven faces in all the above case studies were restricted to the intermediate neighboring faces of the push-pulled faces. We did this in order to be in line with the way of working in Siemens NX for the modeling results of the proposed method to be comparable with those of Siemens NX. However, there are also cases where the user wants driven faces to be broader and include more faces such as second-ring neighboring faces [24]. This basically asks: which $G^1$ connections are to be used in constructing the system of tangent constraints discussed in Section 3.1.2. Although different choices among the $G^1$ connections yield different constraint systems, they make no difference once assembled with Eq. (3), (4), and (5), as long as they are well-constrained. Thus, the choosing scheme being used does not alter the essence of the proposed method in this work. Case study 7 (Fig. 15) shows an example in which different choosing schemes can be incorporated into the proposed method, based on a simplified version of the model used in case study 6. In Fig. 15b, driven faces are intermediate neighboring faces; in Fig. 15c, driven faces include intermediate neighboring faces and second-ring neighboring faces; in Fig. 15d, driven faces are intermediate neighboring faces, and both the face position and face size parameter (i.e., the radius) are made changeable to accommodate the $G^1$ connections.

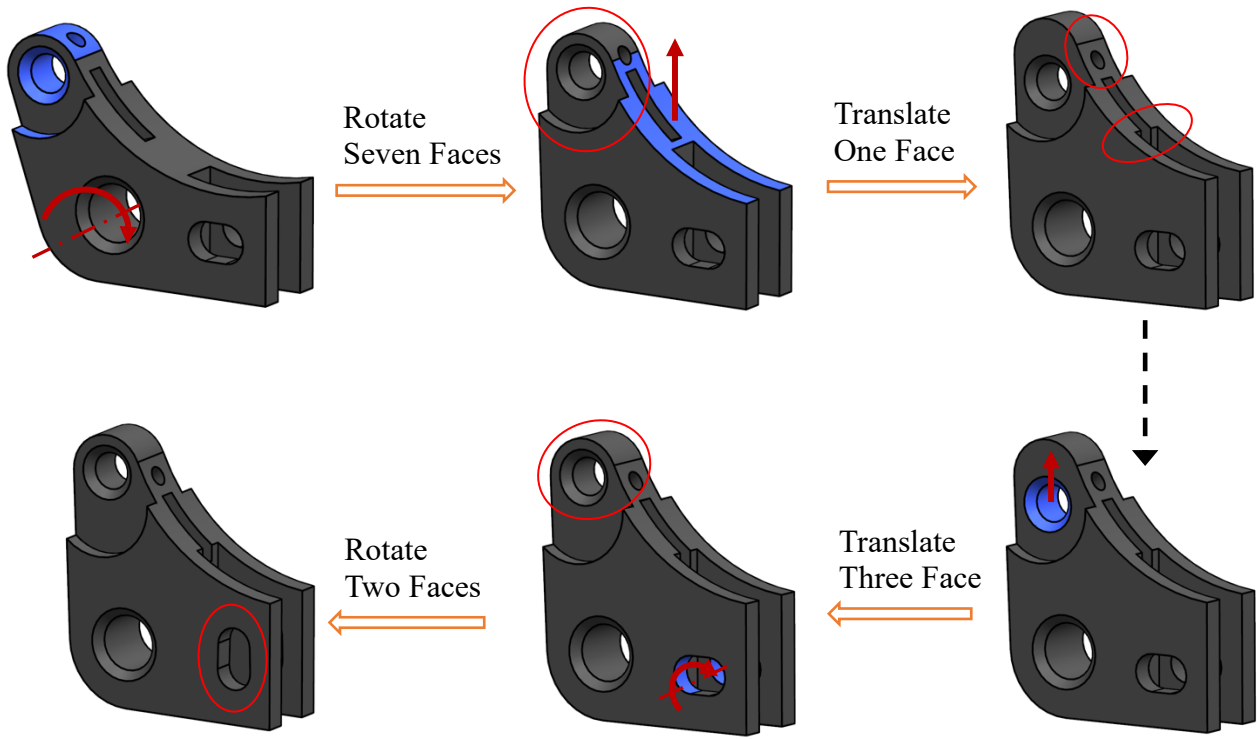

Figure 14: Modeling results of various push-pull edits in a row using the proposed method (circles indicate changed parts).

Although the proposed method can interface with any possible choosing scheme, the present work, in its current form, is not able to automatically determine which of the choosing schemes is to be used, given a specific push-pull edit. The user needs to specify the choosing scheme and uses it as an input to the proposed method. This states the main limitation of the present work. Having an automatic mechanism for decisions among the choosing schemes is absolutely desirable. This is, however, a challenging task, partly because the decision is closely related to the user's design intent. Design intent is generally too complicated to infer satisfactorily by the computer [25].

## 5. Conclusions

A robust method for push-pull direct modeling of quadric B-rep models while preserving smooth connections has been presented in this paper. This kind of direct modeling is of practical significance, but current academic methods and industrial implementations are far from being sufficient in terms of robustness. The major challenge of the robustness issues has been found to be the lack of information on motions of movable boundary faces during push-pull moves and effectiveness towards GTI detection. A novel, effective method has thus been proposed to solve the challenge and attain robust push-pull direct modeling. This method features the ability to detect GTI while avoiding the reliance on the motion information. Case studies and comparisons have been conducted to validate the proposed method.

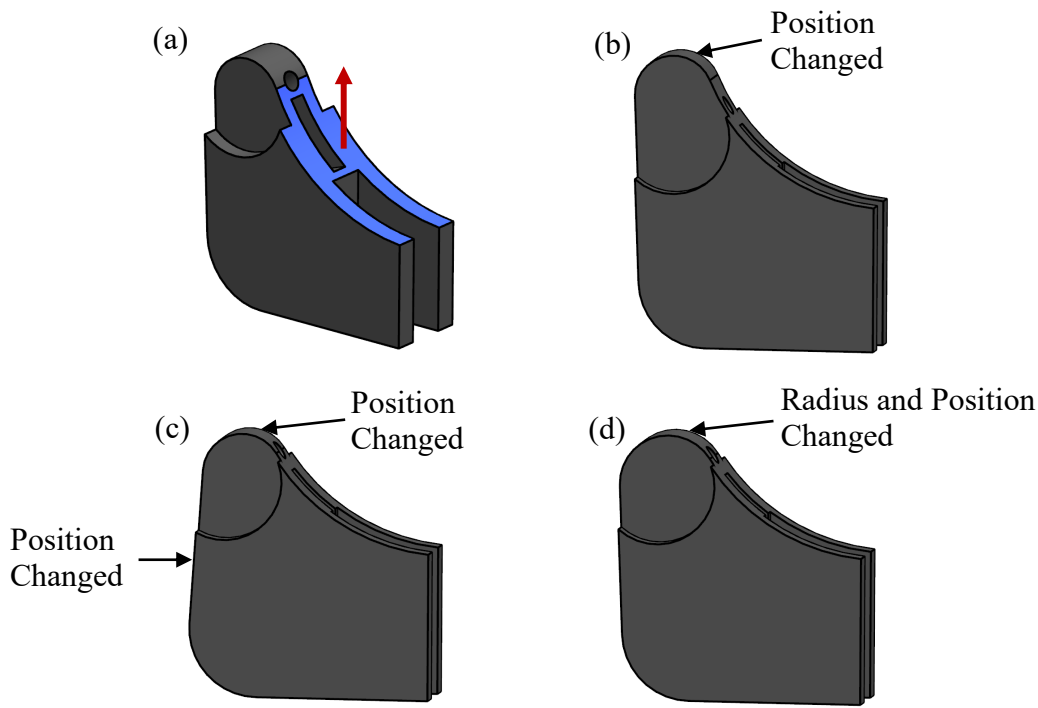

Figure 15: Push-pulling the bracket model under various $G^1$ connection choosing schemes: (a) push-pull edit; (b), (c), and (d) modeling results.

A couple of practical notes need to be made here. In the GTI resolution module (Section 3.2), Boolean operations have been used. Such operations are often criticized for not being very stable. This is true to a certain extent as a systematic, theoretical treatment to this issue is still unavailable. Nevertheless, from a practical perspective, Boolean operations implemented in most geometric modeling kernels have been refined through decades of incremental improvements (in particular, use of the adaptive tolerance technique [26]), and they now work fairly reliably, as reported by experts from industry [27]. It should also be noted that, as the proposed method conducts GTI detection through solving systems of nonlinear equations, the computation load may become significant when the push-pull edit involves many GTI critical points and the model being push-pulled is complex. Geometric constraint system decomposition and parallel computing techniques may be used to address this issue. Improving the proposed method's computational efficiency is among the direct modeling research studies to be carried out in our research group. In addition, combining the present work with virtual/augmented reality techniques is of great interest for future studies.

**Acknowledgment**

This work was in part funded by the Natural Sciences and Engineering Research Council of Canada (NSERC) under the Discovery Grants program.

**References**

[1]   Tornincasa S, Monaco FD. The future and the evolution of CAD. in Proceedings of the 14th International Research/Expert Conference, 2010, p. 11–8.

[2]   Ault H, Phillips A. Direct modeling: easy changes in CAD. in Proceedings of the 70th ASEE Engineering Design Graphics Division Midyear Conference, 2016, p. 99–106.

[3]   Zou Q, Feng H-Y. Push-pull direct modeling of solid CAD models. Advances in Engineering Software 2019;127:59–69.

[4]   Hashemian A, Imani BM. Surface fairness: a quality metric for aesthetic assessment of compliant automotive bodies. Journal of Engineering Design 2018;29:41–64.

[5]   Piegl LA, Tiller W. The NURBS book. Springer; 1997.

[6]   Rabbani T, Van Den Heuvel F. Efficient Hough transform for automatic detection of cylinders in point clouds. in Proceedings of the 11th Annual Conference of the Advanced School for Computing and Imaging, 2005, p. 60–65.

[7]   Requicha AAG. Representations for rigid solids: theory, methods, and systems. ACM Computing Surveys 1980;12:437–64.

[8]   Lipp M, Wonka P, Müller P. PushPull++. ACM Transactions on Graphics 2014;33:1–9.

[9]   Rossignac JR. Issues on feature-based editing and interrogation of solid models. Computers and Graphics 1990;14:149–72.

[10]  Woo Y, Lee SH. Volumetric modification of solid CAD models independent of design features. Advances in Engineering Software 2006;37:826–35.

[11]  Kim BC, Mun DW. Stepwise volume decomposition for the modification of B-rep models. The International Journal of Advanced Manufacturing Technology 2014;75:1393–403.

[12]  Fu J, Chen X, Gao S. Automatic synchronization of a feature model with direct editing based on cellular model. Computer-Aided Design and Applications 2017;14:680–92.

[13]  Hoffmann CM, Kim KJ. Towards valid parametric CAD models. Computer-Aided Design 2001;33:81–90.

[14]  Van der Meiden HA, Bronsvoort WF. Tracking topological changes in parametric models. Computer Aided Geometric Design 2010;27:281–93.

[15]  Hidalgo M, Joan-Arinyo R. Computing parameter ranges in constructive geometric constraint solving: implementation and correctness proof. Computer-Aided Design 2012;44:709–20.

[16]  Bondy JA, Murty USR. Graph Theory with Applications. Macmillan; 1976.

[17]  Braid IC. Geometric modelling. in Advances in Computer Graphics I, Springer; 1986, p. 325–62.

[18]  Do Carmo MP. Differential Geometry of Curves and Surfaces. Prentice Hall; 1976.

[19]  Bettig B, Shah J. Derivation of a standard set of geometric constraints for parametric modeling and data exchange. Computer-Aided Design 2001;33:17–33.

[20]  Zou Q, Feng H-Y. Variational B-rep model analysis for direct modeling using geometric perturbation. Journal of Computational Design and Engineering 2019;6:606–16.

[21]  Zou Q, Feng H-Y. A decision-support method for information inconsistency resolution in direct modeling of CAD models. Advanced Engineering Informatics 2020;44:101087.

[22]  Möbius J, Kobbelt L. OpenFlipper: an open source geometry processing and rendering framework. in Proceedings of the International Conference on Curves and Surfaces, 2012, p. 488–500.

[23]  Hoffmann CM. Robustness in Geometric Computations. Journal of Computing and Information Science in Engineering 2001;1:143–55.

[24]  Botsch, Mario, Leif Kobbelt, Mark Pauly, Pierre Alliez. Polygon mesh processing. CRC press; 2010.

[25]  Camba JD, Contero M. Assessing the impact of geometric design intent annotations on parametric

model alteration activities. Computers in Industry 2015;71:35–45.
[26] Jackson DJ. Boundary representation modelling with local tolerances. in Symposium on Solid Modeling and Applications, 1995, p. 247–53.
[27] Allen G. Geometric Modeling Problems in Industrial CAD/CAM/CAE. in Proceedings of the 10th SIAM Conference on Geometric Design and Computing, 2007, p. 109–26.